\documentclass{aastex}
\slugcomment{Submitted to ApJ Letters on 6/October/1999}
\shorttitle{eta Car binarity}
\shortauthors{Damineli et al.}
\begin{document}
\title{$\eta$ Carinae: Binarity Confirmed}
\author{Augusto Damineli\altaffilmark{1}}
\affil{Instituto Astron\^{o}mico e Geof\'{\i}sico da USP,
Av. Miguel Stefano 4200, 04301-904 S\~{a}o Paulo, Brazil}
\email{damineli@iagusp.usp.br}
\altaffiltext{1}{Based on data collected at LNA (Laborat\'orio Nacional
de
 Astrof\'{\i}sica/CNPq, Brazil) and ESO (European
 Southern Observatory, Chile).}
\author{Andreas Kaufer}
\affil{European Southern Observatory, Alonso de Cordova 3107,
Santiago 19, Chile}
\author{Bernhard Wolf, Otmar Stahl}
\affil{Landessternwarte, K\"onigstuhl, D-69117 Heidelberg,
Federal Republic of Germany}
\and
\author{Dalton F. Lopes and Francisco X. de Ara\'ujo}
\affil{Observat\'orio Nacional,  R. General Jos\'e Cristino 77,
20921-400 Rio de Janeiro, Brazil}

\begin{abstract}
We report the recovery of a spectroscopic event in $\eta$ Carinae
in 1997/98 after a prediction by Damineli (1996).
A true periodicity with $ P=2020\pm5$ days (0.2\% uncertainty)
is obtained. The line intensities and the radial-velocity curve
 display a phase-locked behavior implying that the
energy and dynamics of the event repeat from cycle to
cycle. This rules out S Doradus oscillation or multiple
shell ejection by an unstable star as the explanation
of the spectroscopic events.
A colliding-wind binary scenario is supported by our spectroscopic
data and by X-ray observations. Although deviations from
a simple case exist around periastron, intensive monitoring
during the next event (mid 2003) will be crucial to the
understanding of the system.

\end{abstract}


\section{Introduction}

Eta Carinae is a key object to understand the evolution of massive
stars.
It is one of the most luminous stars known and underwent a
great eruption that created the Homunculus 150 years ago.
Since then, the nature of $\eta$ Carinae has been the subject
of an intense debate. However, the star does not fit any of the
proposed models - see Davidson \& Humphreys (1997) for a review.
In the last three decades the prevailing view was that $\eta$ Car
is a single luminous blue variable star (LBV).
The spectroscopic events - fading of high excitation
lines - occasionally observed, were interpreted
as being imprints of S Doradus oscillations \citep{DSKW}.
The regularity of the 5.5 year cycle, however,
makes such a hypothesis unlikely \citep{AD96}.
The binary system proposed by Damineli, Conti \& Lopes
(1997 - hereafter DCL)
brought new insight to the problem.
The solution to the radial-velocity (RV) curve of
the broad component in the Paschen$-\gamma$
line resulted in two massive stars
($M_{1}\sim$$M_{2}\sim70M_{\odot}$).
The large mass-loss rates of the stars
would imply strong wind-wind collision
producing thermal hard X-rays. The
highly eccentric orbit would produce
phase-locked X-ray variability. Those features
are in accord with the observations \citep{C98}
indicating that $\eta$ Car is a colliding-wind
binary (CWB).
Davidson (1997) presented an
alternative solution to the DCL data, leading to
similar characteristics but larger eccentricity. The
parameters of the binary system, however, were not trustworthy
because they were derived from only one cycle and based on
a single emission line. The prediction of a spectroscopic event
for late 1997 through early 1998 opened an outstanding
opportunity to test this model against the, at the time
dominant, idea of recurrent shell ejection.

The expected event indeed occurred at the right time
as shown by the spectroscopic data analyzed in the next section.
The event was also detected at radio wavelengths
\citep{AD} and in X-rays,
displaying a dramatic variation \citep{I}. A full
2-D hydrodynamic simulation \citep{SP} and even a simple
CWB model \citep{I} reproduce the overall behavior of
$\eta$ Car in X-rays. However, significant discrepancies
remain around periastron. This is not unexpected,
taking into account that near periastron the stars are
close enough for deviations from spherical symmetry
 to be important. The wind is probably
 non-spherical and an equatorial disk may surround
 the primary star.

The event observed in optical/near-infrared spectral lines and
at radio wavelengths is due to a cause different from that in X-rays,
despite their temporal coincidence. Following  Damineli,
Lopes, \& Conti (1999),
the secondary star is the main source of
hard photons, not the wind-wind colliding zone.
The immersion of the secondary star into the
companion's wind prevents hard photons from reaching
 the external regions
of the wind of the primary and the circumstellar gas.
The resulting effect is a temporary drop in the degree
of ionization and, consequently, the fading of the
high excitation lines.

Although preliminary results indicate that
$\eta$ Car is
likely to be a binary, critical tests must be
carried out to provide a definitive proof.
Such tests are provided in this letter,
based on the data collected
in a spectroscopic campaign initiated in 1989 that
covered twice the 5.5 year cycle.

\section{A True Periodicity}

Spectroscopic observations of $\eta$ Car were carried out at
the European Southern  Observatory - ESO/Chile and at the
National Astrophysical Laboratory -
LNA/Brazil - see Wolf et al. (1999) and Damineli et al. (1999)
for details.
The [\mbox{S\,{\sc iii}}] and [\mbox{Ar\,{\sc iii}}]
spectral lines plummeted in November 1997 and
disappeared a month later \citep{LD}. Around that time,
the near-infrared light-curve reached a local maximum
\citep{WL} and the spectroscopic lines a minimum,
similar to those reported by
Whitelock et al.(1994) and by Damineli (1996).
Since we have detailed observations
of the 1992 event, we folded them with the 1998 event.
The sawtooth  structure in the RV curve provides
an accurate measure of the elapsed time between the
last two events. We plotted in Figure 1a the RV variations of the
broad component in \mbox{He\,{\sc i}} $\lambda$6678
for the 1998 event (solid line)
superposed with the previous event shifted by
2020 days (dotted line).
The matching was obtained by shifting one
relatively to the other by $2020\pm5$ days.
The same recurrence time was derived from
Pa-$\gamma$ although the uncertainty is larger for
this line due to a smaller number of observations.

%\placefigure{fig1}

We should not expect, a priori,  that line intensities
would result in the same elapsed time as for
RV. The narrow component in [\mbox{S\,{\sc iii}}] $\lambda$6312
 and  \mbox{He\,{\sc i}} $\lambda$6678 lines
disappeared around 12 December 1997 (JD 2450794$\pm$2 days).
The same was observed on 2 June 1992 (JD 2448775$\pm$5 days),
resulting in a recurrence time $2019\pm7$ days, compatible
with that from RV curve.
This time scale is also in agreement with $P=2014\pm50$ days
proposed by DCL based on the 1948, 1965, 1981,
1987 and 1992 events. This is an indication that
the  previous events resembled closely the last two
ones we have observed: deep and brief. If
the events had lasted longer, no obvious
strict periodicity would have been hinted from such a short
list of occurrences. If line intensity
fadings had not been so remarkable they would have
been mistaken
as secondary fluctuations that are frequent
 in the $\eta$ Car spectrum.

An additional way of checking whether
the periodicity is true is by comparing our data with those
 of Gaviola (1953). The description of the spectrum collected
by that author in 1948 is detailed enough to
show that $\eta$ Car was very close to the center
of a spectroscopic event on April 19th.
He reported that [\mbox{N\,{\sc ii}}] $\lambda$5755 was
 fainter than [\mbox{Fe\,{\sc ii}}]
$\lambda$5748. Our spectra collected during
the last two events show that such an inversion of intensities
 remains for less than three months around the center of an event.
The 9 cycles  between the 1948 and the 1998 events constrain
very well the uncertainty in the periodicity: $P=2019\pm10$ days.
 This figure is in excellent agreement with
that described above, based on the last two events.
We can conclude that the events in $\eta$ Car are truly periodic
with $P=2020\pm5$ days (or $P=5.53\pm0.01$ years).

The equivalent widths of the spectral lines,
especially \mbox{He\,{\sc i}} and the high-excitation
forbidden lines, are quite
 predictable from cycle to cycle.
In Figure 1b we display the
\mbox{He\,{\sc i}} $\lambda$6678 line
during the events of 1998 and 1992 (the
latter one shifted by 2020 days).
 This figure implies that the emitting volume
 and the speed of gas display a phase-locked
 behavior.
Due to the fact that no luminous hot star has
 been observed
pulsating in such a regular fashion and ejecting
identical shells, we rule out
stellar instability as the mechanism
responsible for the
spectroscopic events in $\eta$ Carinae.

\section{The Orbital Parameters}

The orbital solution presented by DCL had
been discussed by Davidson (1997, 1999).
Some concerns have been raised about
the reliability of the broad-line components to
trace the orbital motion.
On the one hand, broad lines are formed inside
 the wind and should display Doppler shifts if the
star were a binary.
On the other hand, radiative-transfer effects could
mask Doppler motion. If the event is
produced by the plunging
of the secondary star into the primary-star wind
(Damineli, Lopes, \& Conti 1999),
the ionization structure of the
wind should change around periastron, mixing
together Doppler and radiative-transfer effects.
Different lines would display
different line-profile variations, making it
difficult to predict what line tracks
the orbital motion better. The best candidates are
lines (broad components)  displaying small
changes in equivalent width and in line
shape, with as faint as possible
P Cygni absorption components.
 Some hydrogen Paschen lines
are free of blends and look suitable for
this purpose, as seen in DCL (their Figure 1).

\mbox{He\,{\sc i}} lines
show strong variability in intensity
and line-profile. Deblending
the broad components
from the nebular and P Cygni contamination
is not as straightforward as in H-Paschen lines.
 The observed changes
 in \mbox{He\,{\sc i}} lines
reflect more the excitation
effects in the wind than Doppler
motion of the stars and this is
why we did not use them in the RV solution.
The  large number of observations we
collected for \mbox{He\,{\sc i}} $\lambda$6678 line,
however, still makes it the best tool
to  measure the  period length.

%\placefigure{fig2}

The RV curve containing all the
available Pa-$\gamma$ and Pa-$\delta$ data
results in the solution displayed
in Figure 2. The best fit
gives: $e=0.75$, $\omega_{1}=275^{o}$,
$T_{periastron}=$JD$2450861$ (17 February 1998),
$K_{1}=50$ km s$^{-1}$ and $fm_{1}=7.5 M_{\odot}$.
Conjunctions occur at phases 0.999 and 0.438.
The standard deviation is $\sigma=11$ km s$^{-1}$.
This result is in general agreement with those of
 Damineli, Lopes, \& Conti (1999) and Davidson (1997).
 The main difference between present
 and previous results is the
smaller mass function, ($fm_{1}$), accommodating a smaller
 mass for the secondary star, which is desirable, since
 its spectrum is not visible in the optical or the UV.
 The mass of the unseen companion is difficult
 to be constrained, as it depends on
 the mass of the primary-star, the mass-loss
 rate and the orbital inclination.
 Regarding the primary-star we adopt DCL's
 M$_{1}\sim70M_{\odot}$, based on the total luminosity
 of the star, the age of Tr16 cluster and standard
 evolutionary models.

Figure 2 gives strong support in favor of binarity
in $\eta$ Car, but it does not necessarily
guarantee that our orbital elements are accurate.
There are two problems with our RV curve. First,
 Pa-$\gamma$ shows changes in the shape of
the broad-line component, indicating that the
wind is not rigidly following the star.
The effect in the RV curve seems to be small
because the $\gamma$ velocity (-12 km s$^{-1}$ )
derived from the RV curve
agrees with V$_{rad}=-7\pm9$ km s$^{-1}$
obtained by Conti, Leep, \& Lorre (1977) from
seven O-type stars in the Carina Nebula.
Second,  data around the periastron
 are scarce, which is a problem for
 a system with such a high eccentricity.
 Acceptable
 eccentricities range from e$\sim$0.65 up to
 e$\sim$0.85, displacing the time of
 periastron by several months
 (up to 0.05 in phase) around
 the center of the spectroscopic event.
 The other orbital elements
 are less sensitive to the details of the
 RV curve around periastron.

\section{Discussion}

The recovery of the 1998 spectroscopic
event within  $<$1\% of the predicted time
supports the hypothesis of true periodicity.
The strict repeatability in the RV curve and line
intensities is unambiguously
in favor of the
binary model and against the idea of an
unstable star as the explanation of the 5.5 year cycle.
Regardless of the particular value adopted
for the orbital
eccentricity (e$>$0.65), the primary star
is a typical LBV and the unseen companion a hotter
and less evolved star.
The orbital parameters
must be regarded as very  provisional, because
data around periastron are scarce and based on
broad emission lines and the physical parameters of
very massive stars are uncertain.
Nevertheless, the success of the CWB models in reproducing
 X-ray luminosity,  temperature, $N_H$, and the
  dip in the light curve, add credibility to the solution.
Fitting  the X-ray light curve by CWB models presents
some problems around periastron, as shown by
 Ishibashi et al. (1999) and Stevens \& Pittard (1999).
However, non-spherical symmetry in the stellar wind
 or a circumstellar disk might remove the disagreement
 between observations and CWB models.

However, the $\eta$ Car wind is not steady and probably not
homogeneous, producing a
 RV curve not as simple as in classical
cases. Significant profile variations
have been seen in some spectral lines around periastron
(Davidson 1999), indicating that gas blobs are thrown away
from the stars, a reminiscence of shell ejection
suggested in previous works.
This time, however, the mechanism is not advocated to explain
the whole 5.5 year cycle behavior, but minor perturbations.

Although the binary scenario accounts for the main
observational characteristics of $\eta$ Carinae,
the nature of the spectroscopic events is not
fully understood. Continuous monitoring of the system,
with a better time sampling than
the previously attained, is needed for the next event
in 2003. An observational campaign starting in May
should extend until late September to determine the
RV curve, to late October to survey the
 dip in the X-ray light-curve, and, through December
 to follow the variability in
 different spectral lines.
 The star is not accessible during night-time from August
 through October, which will preclude ground-based
 optical observations.
High spatial resolution will be crucial to disentangle
the stellar intrinsic variability from the associated
light reflections  through the surrounding nebula.
Far-UV spectra are relevant, as the secondary
(hotter) component of the system may be detectable;
consequently, space-based facilities will be invaluable.
 From the ground, infrared and near infrared will be
 required to perform daylight observations.

\acknowledgments
We thank the ESO and
LNA Observatories for the generous
allocation of observing time.
This work was supported by
FAPESP, FINEP and by the Deutsche
Forschungsgemeinschaft (Wo 296/20-1).
A. D. thanks G. Medina Tamco for
helpful discussions.

\clearpage

\begin{figure}
\plotone{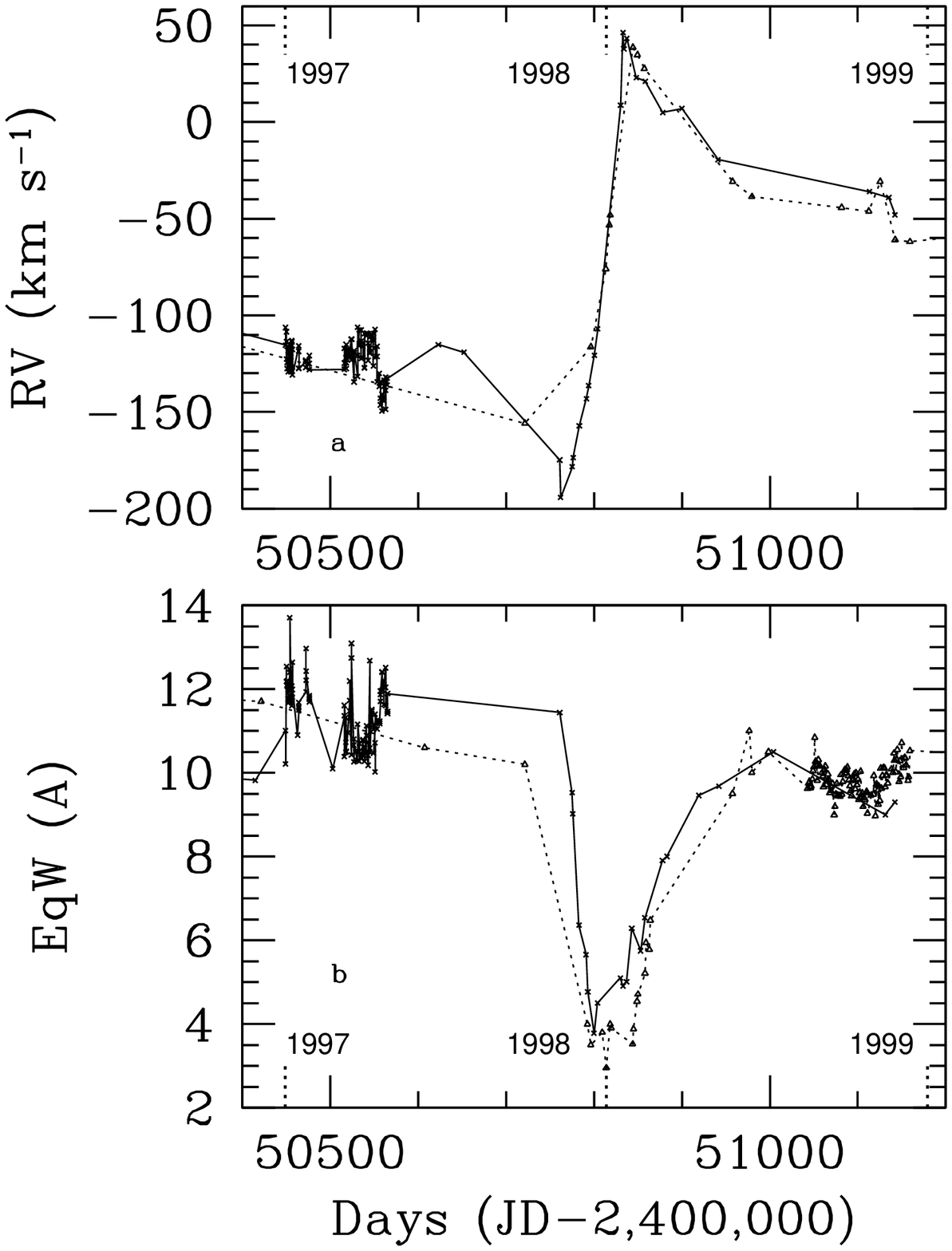}
\caption{Variability during the present cycle (solid line)
folded with the previous one displaced
by 2020 days (dotted line); a) radial velocity (RV) of the broad
component in the \mbox{He\,{\sc i}}
$\lambda$6678 emission  line; b) equivalent width
(EqW) of the blend
\mbox{He\,{\sc i}} $\lambda$6678+[\mbox{Ni\,{\sc ii}}]
$\lambda$6666. Dotted tickmarks indicate
beginning of years. \label{fig1}}
\end{figure}

\begin{figure}
\plotone{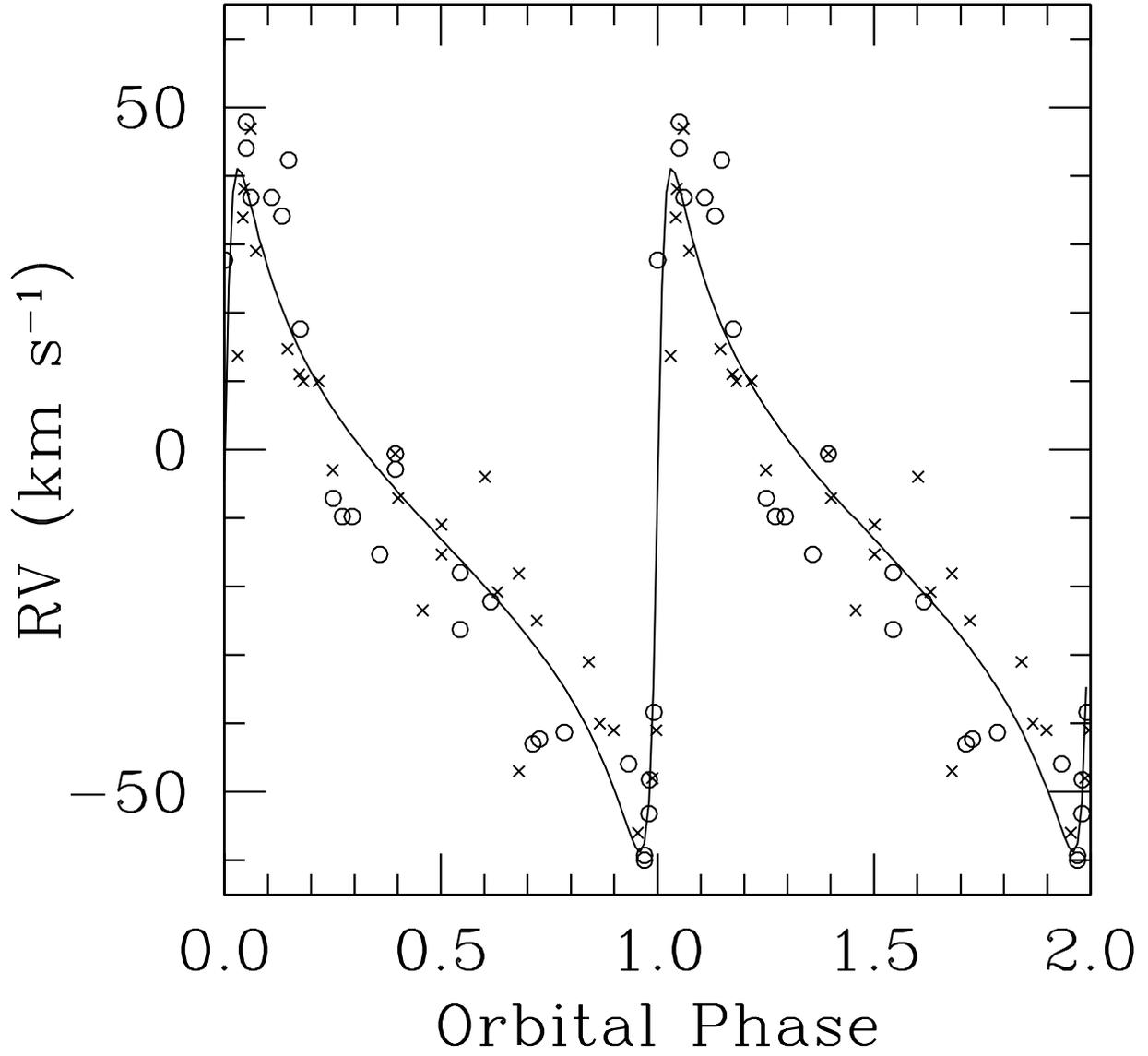}
\caption{Radial velocity curve of Pa$-\gamma$ and Pa$-\delta$
broad-line components. Previous cycle (circles),
present cycle (crosses),
and orbital solution (solid line). \label{fig2}}
\end{figure}

\end{document}